Daniel Hopp

Associate Statistician
Division on
Globalisation and
Development
Strategies, UNCTAD

daniel.hopp@unctad.org

# Economic Nowcasting with Long Short-Term Memory Artificial Neural Networks (LSTM)

## Abstract

Artificial neural networks (ANNs) have been the catalyst to numerous advances in a variety of fields and disciplines in recent years. Their impact on economics, however, has been comparatively muted. One type of ANN, the long short-term memory network (LSTM), is particularly well-suited to deal with economic time-series. Here, the architecture's performance and characteristics are evaluated in comparison with the dynamic factor model (DFM), currently a popular choice in the field of economic nowcasting. LSTMs are found to produce superior results to DFMs in the nowcasting of three separate variables; global merchandise export values and volumes, and global services exports. Further advantages include their ability to handle large numbers of input features in a variety of time frequencies. A disadvantage is the inability to ascribe contributions of input features to model outputs, common to all ANNs. In order to facilitate continued applied research of the methodology by avoiding the need for any knowledge of deep-learning libraries, an accompanying Python library was developed using PyTorch: https://pypi.org/project/nowcast-lstm/.

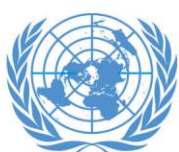

UNITED NATIONS

## Contents

## Acknowledgements

I would like to thank Fernando Cantú, Anu Peltola, Onno Hoffmeister, Henri Luomaranta and Stephen MacFeely for their valuable comments and support.

# 1. Introduction

A defining feature of the 21st century so far has been the explosion in both the volumes and varieties of data generated and stored (Domo, 2017). Almost every industry and aspect of life has been affected by this "data revolution" (Einav and Levin, 2014), (MacFeely, 2020). Simultaneously, rapid advancements in machine learning methods have been made, spurred on in part by the need for novel methods to analyze these new data quantities. Perhaps no methodology has gained greater prominence than the artificial neural network (ANN). ANNs are the engine behind tremendous leaps in fields as disparate as machine translation, image recognition, recommendation engines and even self-driving vehicles. Yet to date, their impact in the field of economic policy has been largely muted or exploratory in nature (Falat and Pancikova, 2015).

This is not to suggest that economic data have been immune to the transformative forces of the data revolution. Quite the opposite in fact, as classical economic data series from national statistical offices (NSO) and other organizations can now be fortified by alternative data sources like never before, helping to provide glimpses into the developments of the global economy with unparalleled granularity and timeliness (Glaeser et al., 2017). The COVID-19 pandemic and ensuing economic crisis showcased this, with analysts and policy-makers gaining insight to the rapidly evolving economic situation from such alternative data sources as Google mobility data (Yilmazkuday, 2021), booking information from dining apps (OpenTable, 2021) and transaction data from e-commerce sites (Statista, 2021), among many others.

The availability of a broad range of novel, timely indicators should ostensibly have led to significant advances in the field of economic nowcasting, where real-time macroeconomic variables that may be published with a significant lag are estimated based on an array of more timely indicators (Banbura et al., 2010), (Giannone et al., 2008). In reality, the field has not experienced the degree of progress seen in other fields, such as image recognition, in the past 10 years. A large factor in this relative stagnation is the fact that many of the issues facing nowcasting are not addressed by more data alone. Issues such as multicollinearity, missing data, mixed-frequency data and varying publication dates are sometimes even exacerbated by the addition of variables (Porshakov et al., 2016). As such, advancements in the field come from a combination of both new data and methodological developments. Dynamic factor models (DFM) in particular have been found to address many of the data issues inherent in nowcasting (Stock and Watson, 2002), and have been applied successfully in applications such as nowcasting economic growth in 32 countries (Matheson, 2011), nowcasting German economic activity (Marcellino and Schumacher, 2010) and nowcasting Canadian GDP growth (Chernis and Sekkel, 2017). The basic premise of DFMs is that one or more latent factors dictates the movement of many different variables, each with an idiosyncratic component in relation to the factor(s). With historical data, the factor(s) can be estimated from the variables. Subsequently, even in future periods where not all data are complete, the factor(s) can still be estimated and used to generate forecasts for variables that are not yet published, as each variable's relation to the factor(s) has already been estimated.

Despite DFMs' strengths in addressing a wide swath of nowcasting's data issues, the impressive performance of ANNs in other domains raises the question of their performance in nowcasting. ANNs have been applied to economic nowcasting in the past (Loermann and Maas, 2019). However, due to the time-series nature of many economic nowcasting applications, the long short-term memory (LSTM) architecture is better suited to the problem than the traditional feedforward architecture explored in Loermann and Maas (2019). LSTMs are an extension of recurrent neural network (RNN) architecture, which introduces a temporal component to ANNs. LSTMs have been used to nowcast meteorological events (Shi et al., 2015) as well as GDP (Kurihara and Fukushima, 2019).

However, use of LSTMs in nowcasting economic variables remains in its infancy, perhaps partly due to high barriers to their implementation. Many common deep learning frameworks, including Keras and PyTorch, include provisions for LSTMs. However, the implementations are general and require knowledge of the frameworks to successfully implement. As such, a Python library focused on economic nowcasting has been published alongside this paper, available for install on PyPi: https://pypi.org/project/nowcast-lstm/. Hopefully, a more accessible library will help stimulate interest and expand the applications of these powerful models.

The remainder of this paper is structured as follows: the next section will further explain nowcasting and its challenges; section three will explore LSTMs in more detail; section four will examine the LSTM's empirical performance compared with DFMs in nowcasting three series: global merchandise trade exports expressed in both values and volumes and global services exports; section five will introduce and explain the accompanying Python library; the final section will conclude and examine areas of future research.

# 2. Exposition of nowcasting problem

Nowcasting, a portmanteau of "now" and "forecast", is the estimation of the current, or near to it either forwards or backwards in time, state of a target variable using information that is available in a timelier manner. Keith Browning coined the term in 1981 (WMO, 2017) to describe forecasting the weather in the very near future based on its current state. The concept and term remained in the meteorological domain for years before being adopted into the economic literature in the 2000s. The concept of real-time estimates of the macroeconomic situation predates the adoption of the nowcasting terminology, as evidenced by Mariano and Murasawa (2003). However, Giannone et al. (2005) explicitly referenced the term "nowcasting" in its title and the term became commonplace in subsequent years, being applied for example to Portuguese GDP in 2007 (Morgado et al., 2007) and to Euro area economic activity in 2009 (Giannone et al., 2009). The 2010s saw a wealth of papers examining the topic both for a range of target variables as well as with a range of methodologies and models. Targets most often included GDP (Rossiter, 2010), (Bok et al., 2018), and trade (Cantú, 2018), (Guichard and Rusticelli, 2011). Common methodologies include dynamic factor models (DFM) (Guichard and Rusticelli, 2011),  (Antolin-Diaz et al., 2020), mixed data sampling (MIDAS) (Kuzin et al., 2009), (Marcellino and Schumacher, 2010) and mixed-frequency vector autoregression (VAR) (Kuzin et al., 2009), among others. Nowcasting also has relevance in the context of the 2030 Agenda for Sustainable Development (UN, 2015). Many indicators face issues in terms of data quality, availability, timeliness, or all three. As such, nowcasting is being discussed as a possible method of ensuring maximum coverage in terms of indicators (UNSD, 2020).

Economic nowcasting is generally confronted with three main issues regarding data. The first is mixed frequency data, or when all independent variables and the dependent variable are not recorded with the same periodicity. This occurs frequently in economic data, for instance when trying to nowcast a quarterly target variable, such as GDP growth, using monthly indicators. Or estimating a yearly target variable with a mixture of monthly and quarterly variables. The second is the heterogeneous publication schedules of independent variables, frequently referred to as "ragged-edges". Any nowcasting methodology must provide provisions for incomplete or partially complete data, as varying availability of latest data is the reality of most datasets of economic series. Finally, there is the issue of the "curse of dimensionality", which renders many classical econometric methods less effective in the nowcasting context and hinders the application of "big data" to the field (Buono et al., 2017). The problem stems from the nature of many economic variables, where they may have few observations relative to the potential pool of explanatory variables or features. The quarterly target series for UNCTAD's own nowcasts for global merchandise trade, for instance, only began in 2005 (Cantú, 2018). That leaves only 60 observations for training a model at the end of 2020. Meanwhile, many more than 60 potential independent variables can be conceived of to estimate a model of global merchandise trade.

The nowcasting methodologies previously mentioned address these problems in varying ways to achieve better predictions, and LSTMs are no different. The following section will provide background information on their network architecture as well as the characteristics that allow them to address the aforementioned nowcasting data problems.

# 3. ANN and LSTM models

## 3.1 ANNs and RNNs

ANNs are made up of various inter-connected layers composed of groups of nodes or neurons. The structure's conceptual similarity to the biological sort is the source of their name. Each of these nodes receives inputs either from the external data source, the "input layer", or from previous layers, "hidden" and "output" layers, the latter if the final output of the model. The output of a node is found by taking the weighted sum of all its inputs, the connections between individual nodes being the weights, and then running it through a non-linear activation function. In training, these weights are initially randomized, and when the data has passed through all layers of the network, an output is obtained, which is then run through a predefined cost function to assess performance. Then, using calculated gradients, or derivatives of the cost function, the network adjusts its weights to obtain an output with a smaller error, and the process is repeated.

This is of course an oversimplification of the process, however, there exists a vast literature outlining and explaining the methodology of ANNs for those desiring a deeper examination of their mathematics. Those interested can see Sazli (2006), Singh and Prajneshu (2008), or even Loermann and Maas (2019) for an explanation in the nowcasting context.

Traditional feedforward ANNs have a long history of use in time series forecasting (Kohzadi et al., 1996). These models, however, lack an explicit temporal aspect. This can be introduced to their architecture, resulting in recurrent neural networks (RNN) (Amidi and Amidi, 2019). As opposed to the unidirectional relationship between inputs and outputs in feedforward networks, RNNs introduce a feedback loop, where layer outputs can be fed back into the network (Stratos, 2020), (Dematos et al., 1996). This architecture makes RNNs well-suited to applications with a temporal aspect or flow, such as natural language processing or speech processing. However, due to vanishing gradients, RNNs tend to have a very "short" memory, limiting their usefulness in the nowcasting application (Grosse, 2017).

## 3.2 LSTMs

LSTMs introduce a memory cell and three gates: an input, output and forget gate (Chung et al., 2014). Crucially, this architecture then allows gradients to flow unchanged through the network, mitigating the vanishing gradient problem of RNNs and rendering them more suitable for application to the nowcasting problem. Data input to the LSTM network has the shape of *number of observations x number of timesteps x number of features*. The addition of the timesteps dimension allows the model to be trained on multiple lags of each variable, rather than just cotemporaneous observations.

LSTMs' ability to address the first common nowcasting data issue, mixed frequency data, stems from ANNs' ability to learn complex, non-linear relationships in data, a product of multiple neuron layers coupled with non-linear activation functions. More information on activation functions and their role in ANNs can be found in Sharma et al. (2020). As such, mixed frequency data can be fed to the network in the highest frequency available, with lower frequency data having missings at time periods where data are not published. These missing data can then be filled using a variety of approaches, including with the mean, the median, with values sampled from a distribution (Ennett et al., 2001), or with other more complex methods (Smieja et al., 2019). In the analysis performed in this paper, mean replacement was chosen and implemented in the accompanying Python library due to simplicity and empirical performance, as the network learns to recognize these in-between values as containing no novel information.

LSTMs are able to address the ragged-edges problem through no special mechanism other than standard missing-filling methods. These include using ARMA or VAR models to fill in ragged-edges (Kozlov et al., 2018), as well as using the mean or Kalman filters (Doz et al., 2011). The method chosen in the context of LSTM nowcasting can be considered a hyper-parameter to be tuned and tested empirically. At the time of writing, the Python library supports ARMA filling and any n-to-1 series transformation, e.g., mean, median, etc. ARMA filling was used in the analysis performed in this paper due to superior empirical performance compared with other methods.

The last major problem of nowcasting, the curse of dimensionality, is partially addressed by LSTMs' efficiency compared with other methods, i.e. their computation time scales very slowly with the number of variables (Hochreiter and Schmidhuber, 1997), as evidenced by Figure 1.

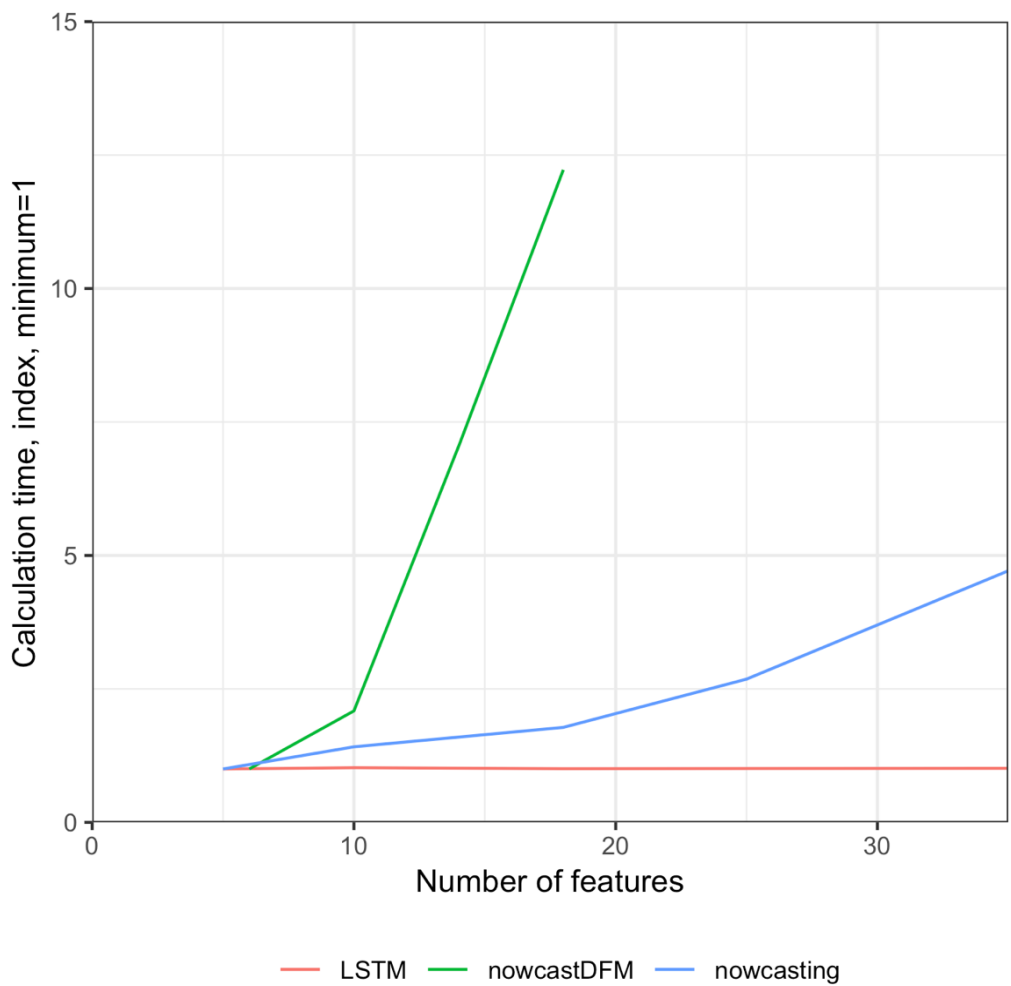

Figure 1. Development of model calculation time depending on number of features

The figure illustrates the development in computation time as more features are added for the LSTM model and two R implementations of the DFM; *nowcastDFM* (Hopp and Cantú, 2020) and *nowcasting* (Marcolino de Mattos, 2019). The DFMs' computation time scales exponentially while the LSTM's time remains nearly constant. The DFM and LSTM both scale linearly with the number of observations, however. The target variable was global merchandise exports in values and the various independent variables were a sample from the same pool presented in section 4. However, for this illustrative case, it is the computation times that are of interest, which display the same general patterns independent of the specific values of the input data.

As a result of this efficiency, a functional model can be trained with many more features than a DFM. Additionally, while standard methods of feature reduction such as principal component analysis (PCA) or Lasso can still be used on a data set intended for use with LSTM networks, their necessity is reduced due to ANNs' robustness to multicollinearity (De Veaux and Ungar, 1994).

Within the LSTM architecture, as in any ANN, there are many choices to be made regarding network architecture and hyper parameters. Some examples include the number of hidden states, the number of layers, the loss function and the optimizer, among many others. The logic for the defaults chosen for the Python library will be discussed in section 5.

# 4. Empirical analysis

## 4.1 Description of data and models

In order to assess the relative performance of LSTMs vs DFMs, three target variables were used: global merchandise exports in both value (WTO, 2020) and volume (UNCTAD, 2020a), and global services trade (UNCTAD, 2020a). These are the same series UNCTAD currently produces nowcasts for using DFMs (UNCTAD, 2020b: 20) and which were examined in a previous UNCTAD research paper (Cantú, 2018). The target series are all quarterly. A large pool of 116 mixed-frequency monthly and quarterly independent series was used to estimate each of the target series. These series are listed in Appendix 1, while more information on any individual series is available upon request. All series were converted to seasonally adjusted growth rates using the US Census Bureau's X13-ARIMA-SEATS methodology (USCB, 2017).

The DFM model used was the same examined in Cantú (2018) and currently in use by UNCTAD. In this model, the DFM is modeled in a state-space representation where it is assumed that the target and independent variables share a common factor as well as individual idiosyncratic components. The Kalman filter is then applied and maximum likelihood estimates of the parameters obtained. This is a common method of estimating DFMs and is explained in further detail in Bańbura and Rünstler (2011). The LSTM model used was that present in the *nowcast_lstm* Python library, which is further explained in section 5, using the average of 10 networks' output with basic hyper-parameter tuning of the number of training episodes or epochs, batch size, number of hidden states, and number of layers. The logic of averaging the output of more than one network to obtain predictions is discussed further in section 5, but see Stock and Watson (2004) for a discussion of forecast combination.

## 4.2 Modelling steps

Hyper-parameter tuning of the LSTM and model performance was evaluated using a training set dating from the second quarter of 2005 to the third quarter of 2016. The test set dated from the fourth quarter of 2016 to the fourth quarter of 2019.

A pool of independent variables was used to ensure the robustness of results, as either model could perform better on a single set of features due to chance. As such, the models' performance was evaluated by taking random samples of between five and 20 features, then fitting both an LSTM and DFM model on this same sample. Both methods' performance was then evaluated on the test set via mean absolute error (MAE) and root-mean-square error (RMSE) on five different data vintages, repeating the process 100 times for each of the three target variables. In this manner, a distribution of relative performance over a wide breadth of independent variables could be obtained. The number of features was restricted to a maximum of 20 due to the high computational time of estimating DFMs with more than this number.

Data vintages in this case refer to the artificial withholding of data to simulate what the availability of data would have looked like at different points in the past. This is important in evaluating model performance in the nowcasting context, as in real life series have varying publication schedules which nowcasting models must be robust to. The five vintages simulated were: two months before the target period, e.g. if the target was the second quarter of 2019, the data as it would have appeared in April 2019; one month before; the month of; a month afterwards; and two months afterwards. The model continues to be evaluated even after the target period has theoretically "passed" as data continue to be published for a given month well after it has passed, depending on the series' individual publication schedule. For example, two months after the second quarter of 2019 simulates being in August 2019, when much more data on the second quarter is available. The variables' publication lags were obtained based on empirical observations from the period from April to November 2020.

## 4.3 Relative performance

Figure 2. LSTM error as a proportion of DFM error

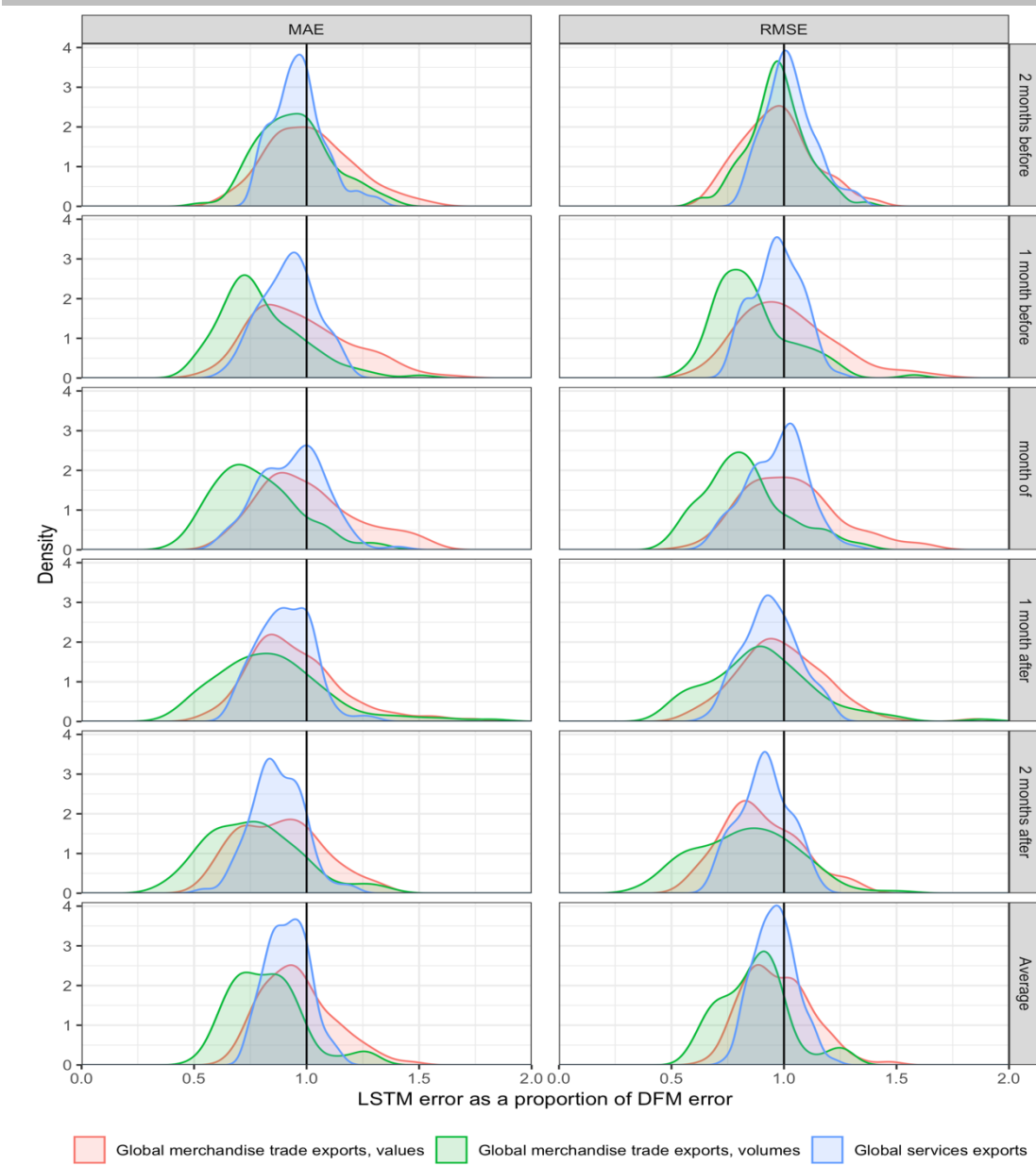

Figure 2 shows the distribution of the LSTM's error as a proportion of the DFM's for each target variable. A value less than one for an individual model indicates better performance on the test set for the LSTM, while a value greater than one indicates worse performance. Consequently, a distribution centered around one, i.e. the vertical line, indicates comparable performance between the two models, while one to the left of the vertical line indicates better performance on average for the LSTM model.

The results clearly favor the LSTM model, obtaining better average performance for both performance metrics across all data vintages and target variables, with the sole exception of RMSE for the two months before services exports vintage. Tables 1, 2 and 3 display the average performance metrics for the two models over the sample of 100 different feature combinations, as well as the results using a simple autoregressive model as a benchmark. A one-tailed t-test was performed on the LSTM and DFM errors to ascertain the significance of these differences in performance, with the alternative hypothesis that the LSTM errors were lower. Results are displayed in the LSTM columns.

### Table 1. Average performance metrics, global merchandise trade exports, values

| Vintage | ARMA MAE | LSTM MAE | DFM MAE | ARMA RMSE | LSTM RMSE | DFM RMSE |
|---|---|---|---|---|---|---|
| 2 months before | 0.0177 | 0.0149 | 0.0150 | 0.0233 | 0.0176** | 0.0185 |
| 1 month before | 0.0177 | 0.0112* | 0.0117 | 0.0233 | 0.014 | 0.0141 |
| month of | 0.0177 | 0.0115 | 0.0118 | 0.0233 | 0.0142 | 0.0142 |
| 1 month after | 0.0168 | 0.0108*** | 0.0117 | 0.0217 | 0.0138 | 0.0142 |
| 2 months after | 0.0168 | 0.0094*** | 0.0109 | 0.0217 | 0.0119*** | 0.0135 |
| Average | 0.0173 | 0.0115** | 0.0122 | 0.0227 | 0.0143* | 0.0149 |

*Note: *p<.05   **p<.01   ***p<.001*

### Table 2. Average performance metrics, global merchandise trade exports, volumes

| Vintage | ARMA MAE | LSTM MAE | DFM MAE | ARMA RMSE | LSTM RMSE | DFM RMSE |
|---|---|---|---|---|---|---|
| 2 months before | 0.0085 | 0.006** | 0.0064 | 0.0097 | 0.0075** | 0.0078 |
| 1 month before | 0.0085 | 0.0051*** | 0.0066 | 0.0097 | 0.0066*** | 0.0079 |
| month of | 0.0085 | 0.0049*** | 0.0065 | 0.0097 | 0.0063*** | 0.0079 |
| 1 month after | 0.0084 | 0.0045*** | 0.0057 | 0.0108 | 0.0059*** | 0.0069 |
| 2 months after | 0.0084 | 0.0042*** | 0.0056 | 0.0108 | 0.0054*** | 0.0067 |
| Average | 0.0085 | 0.0049*** | 0.0062 | 0.0101 | 0.0063*** | 0.0074 |

*Note: *p<.05   **p<.01   ***p<.001*

### Table 3. Average performance metrics, global services exports

| Vintage | ARMA MAE | LSTM MAE | DFM MAE | ARMA RMSE | LSTM RMSE | DFM RMSE |
|---|---|---|---|---|---|---|
| 2 months before | 0.0119 | 0.0123*** | 0.0129 | 0.0151 | 0.0154 | 0.0152 |
| 1 month before | 0.0119 | 0.0103*** | 0.0113 | 0.0151 | 0.0135** | 0.0140 |
| month of | 0.0119 | 0.0103*** | 0.0111 | 0.0151 | 0.0135** | 0.0141 |
| 1 month after | 0.0119 | 0.0103*** | 0.0115 | 0.0151 | 0.0137*** | 0.0146 |
| 2 months after | 0.0119 | 0.0101*** | 0.0117 | 0.0151 | 0.0134*** | 0.0147 |
| Average | 0.0119 | 0.0107*** | 0.0117 | 0.0151 | 0.0139*** | 0.0145 |

*Note: *p<.05   **p<.01   ***p<.001*

## 4.4 Comparison with DFM

The fact that the LSTM performed better than the DFM on average for all three target variables across almost all vintages and both performance metrics is strong evidence for their relevance in the economic nowcasting space. Of course, it does not indicate that LSTMs are superior to DFMs in every instance. The results rather provide some evidence that they can be a competitive alternative to DFMs and have the potential to become a more commonly used methodology in nowcasting.

It is worth mentioning, however, that due to the time and computational constraints inherent in the evaluation of 300 separate models, the hyper-parameter tuning performed for the LSTM models was scaled back relative to what would normally be performed for a single production model. As such, it is reasonable to expect that with finer grid search, the LSTM would have been able to obtain even better results. There are, however, characteristics of the methodology with pros and cons relative to DFMs that are independent of predictive performance.

One of the pros relative to the DFM was discussed in section 3 and illustrated in Figure 1. Namely, LSTMs' ability to handle many more features than the DFM before coming up against computational bottlenecks. This could be beneficial by lessening the need for variable selection in the early stages of an analysis, easing the obtainment of initial results. Additionally, a model is able to be reliably trained on any given set of features and values, which is not the case for the DFM, the training of which may fail if input matrices are non-invertible.

A third advantage is the ability to easily use mixed-frequency variables with no corresponding change in the underlying modeling and formulas. Annual, quarterly, monthly, and even theoretically daily data can be combined in a single model just by changing the structure or frequency of the input data, as explained in section 3.2.

Computational speed is more difficult to ascribe to either method as an advantage. There are many factors affecting the computation time of the two models. For DFMs, this includes the number of features, the number of observations, and especially the time taken for maximum likelihood convergence. For LSTMs, this includes the number of observations, as well as nearly all of the hyper-parameters. As such, there are cases where either method can be faster. Even still, training a single LSTM network regardless of hyper-parameters is usually faster than estimating a DFM on the same data. For instance, in the 300 model runs of this analysis, this was the case 97 per cent of the time, with the LSTM taking on average just 22 per cent of the time needed to estimate the DFM. However, the results in Figure 2 were obtained by fitting 10 LSTM models and averaging the result, in which case the LSTM was faster just 46 per cent of the time, taking on average 2.16 times as long to estimate compared with the DFM. These numbers are slightly skewed in favor of the DFM however, as the number of features was restricted to a maximum of 20. Models with a number of features above this would favor the LSTM in computation time.

The fact that results were evaluated using 10 networks for the LSTM has to do with one of their disadvantages relative to DFMs, namely, the stochastic nature of ANNs. Ten LSTM networks trained on the same data will output ten different predictions due to the randomization of initial weights, which is not the case for DFMs. Training many networks and taking their average predictions is a way to mitigate this characteristic. Figure 3 illustrates how the distribution of predictions develops as more networks are used.

Figure 3. Distribution of predictions of the same target observation

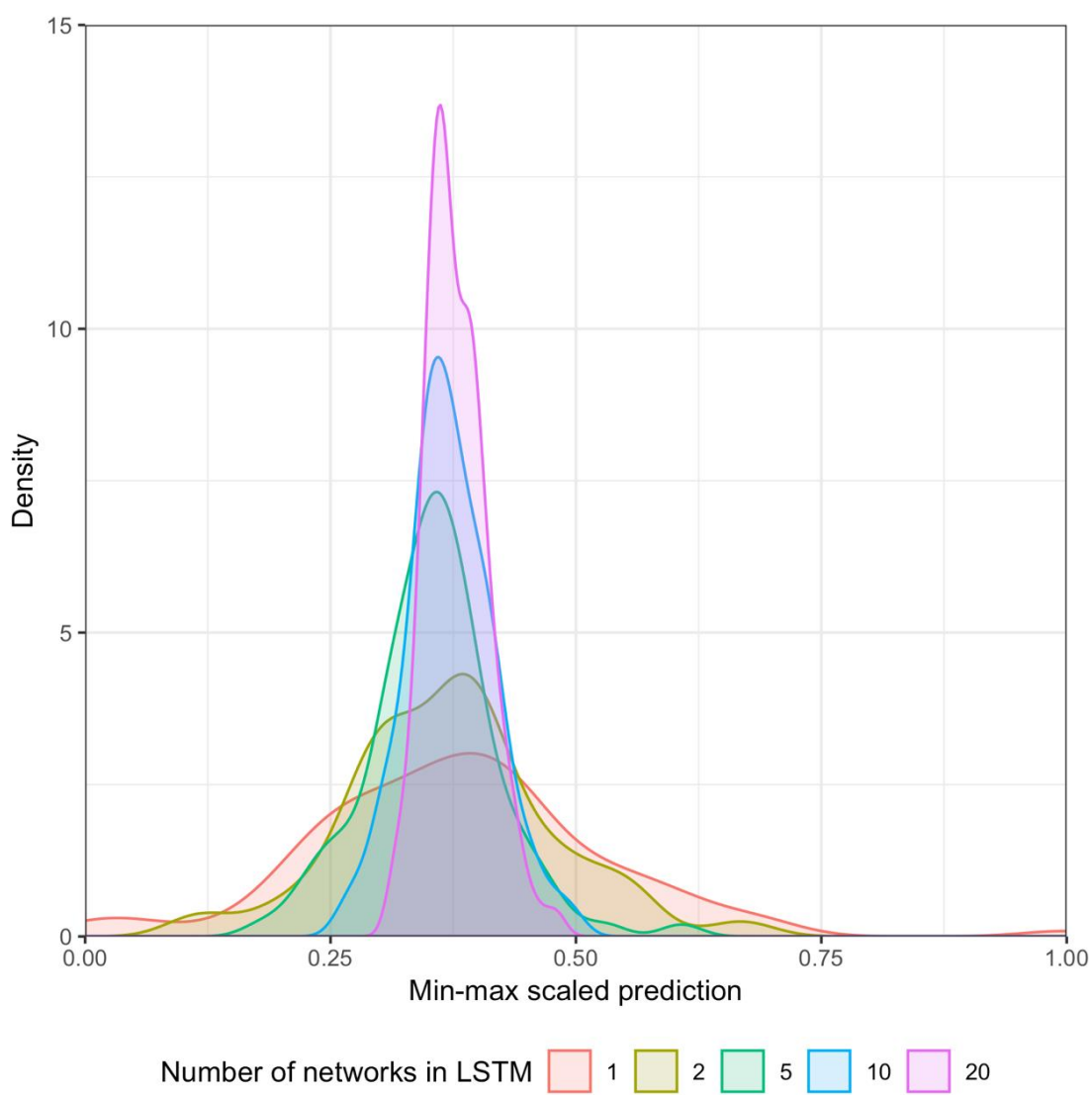

The distributions were obtained by taking a single model predicting global merchandise exports values and training an LSTM model with between one and 20 networks. This was repeated 100 times, thus generating 100 predictions for a single target period, creating the distributions. Variance decreases as more networks are added. While adding more networks can mitigate the stochastic nature of LSTMs' predictions, adding very many can substantially increase computation time while never achieving perfectly consistent outputs.

A final disadvantage LSTMs have compared with DFMs is the lack of interpretability in their parameters, and the consequent lack of inference as to what is driving changes in the model. DFMs have the advantage of being able to offer precise insights to various features' impact on predictions, as illustrated by the New York Fed's nowcasts (Federal Reserve Bank of New York, 2021). This is a well-known characteristic of ANNs in general (Fan et al., 2020). In this regard, there is opportunity for further research into applying existing ANN interpretability methods, such as activation maximization or sensitivity analysis (Montavon et al., 2018), to the nowcasting LSTM framework, but it is outside the scope of this paper.

# 5. Python library

Despite LSTMs' encouraging performance in the nowcasting context, there does not currently exist an easy-to-use implementation in either R or Python that doesn't require a deeper knowledge of neural networks in general or of popular deep learning frameworks like PyTorch or TensorFlow. Therefore, a Python library using PyTorch was developed in the course of making the analysis presented in this paper: *nowcast_lstm,* available for installation on PyPi: https://pypi.org/project/nowcast-lstm/ (Hopp, 2020). The git repository contains instructions and an example Jupyter Notebook, but this section will outline the library's usage and functionality briefly.

**Install and data setup**

The library is hosted on PyPi and can be installed with the terminal command:

```
pip install nowcast-lstm
```

More information on installing Python packages can be found here: https://packaging.python.org/tutorials/installing-packages/. As input, the model expects a Pandas DataFrame with a column for the date (in datetime/timestamp format), columns for features, and a column for the target variable. Rows should be in the frequency of the most granular feature. For example, a mix of monthly and quarterly features should have a row for each month, with quarterly features having values only every three months. The in-between values should be encoded as Numpy nan's. Ideally, the data should be stationary and seasonally adjusted. The model may be able to generate sensible results if this is not the case, but its performance in this scenario has not been validated.

**Usage**

Instantiating and training a model and inference is illustrated below, assuming *data* to be a Pandas DataFrame adhering to the conditions laid out in the previous section, and *test_data* to be a Pandas DataFrame with the same columns but of a different time period:

```
from nowcast_lstm.LSTM import LSTM
model = LSTM(data, "target_col_name", n_timesteps=12)
model.train()
model.predict(test_data)
```

The above code will instantiate and train a model with the default parameters, then generate predictions on a new dataset. Full information on all parameters can be found in the README (Hopp, 2020). Default parameters and model architecture were found by performing grid search hyper-parameter tuning on many samples of random features for each of the three target variables discussed in section 4, then selecting parameters that had the best RMSE over all three targets. While the default parameters generate good results for a wide variety of input data and can be suitable for an initial analysis, hyper-parameter tuning should eventually be performed to ensure the best possible results for the project-specific data are obtained.

A trained model can be saved and loaded using the *dill* library, code below:

```
import dill

# saving a trained model
dill.dump(model, open("trained_model.pkl", mode='wb'))

# loading a trained model
trained_model = dill.load(open("trained_model.pkl", "rb", -1))
```

Further information on the library, including topics such as the filling of missing values, information on hyper-parameters, artificial generation of ragged datasets, and properties of the *LSTM* object are available in the README and example Jupyter Notebook (Hopp, 2020).

# 6. Conclusion

The events of 2020 have shown the value of timely, accurate estimates of macroeconomic series to help inform policy decisions. This paper provides evidence for stronger consideration of LSTMs for this purpose, as well as introduces a Python library to facilitate future research. LSTMs were shown to produce superior predictions compared with DFMs on three different target series: global merchandise trade exports expressed in both values and volumes and global services exports, and over five different data vintages.

In addition to better empirical performance for the three target series, LSTMs provide advantages over DFMs by being able to handle large numbers of features without computational bottlenecks, not relying on the invertibility of any matrices, thus being able to be fit on any dataset, and the ability to use any mixture of frequencies in features or target. Disadvantages relative to DFMs include LSTMs' stochastic nature, the lack of interpretability in their coefficients, and opacity regarding feature contribution to predictions.

The *nowcast_lstm* library can facilitate the use of LSTMs in economic nowcasting by lowering the barrier to experimentation. LSTMs' ability to reliably generate predictions on a large number of input features makes it easier to quickly verify whether or not a given series has the potential to be nowcast, a characteristic that could help expand the variety and quantity of economic variables monitored via nowcasting.

There remains much scope for future research and development on this topic. Further testing should be performed to verify LSTMs' performance on a wider variety of series and frequency mixtures. More hyper-parameter tuning could be performed to see if tweaking other aspects of model architecture could result in even better results. There is also much scope for exploring different methods of filling missing values beyond ARMA or mean-filling. Finally, methods for interpreting LSTMs and ascertaining feature contribution to predictions would increase the method's viability as a policy-informing instrument. The library could then be extended in the future to incorporate any improvements to performance or functionality deriving from future research, continuing to facilitate the adoption and development of the methodology in the nowcasting domain.

# Appendix

| Variable | Frequency | Source |
| --- | --- | --- |
| Exports of services, Japan | monthly | BOJ |
| Export volumes, Africa and Middle East | monthly | CPB |
| Export volumes, China | monthly | CPB |
| Export volumes, Eastern Europe and CIS | monthly | CPB |
| Export volumes, emerging Asia | monthly | CPB |
| Export volumes, Euro area | monthly | CPB |
| Export volumes, Japan | monthly | CPB |
| Export volumes, United States | monthly | CPB |
| Export volumes, world | monthly | CPB |
| Unit value of exports | monthly | CPB |
| Industrial new orders, Euro area | monthly | ECB |
| Harpex charter rate index | monthly | Eikon |
| Manufacturing purchasing managers' index, United States | monthly | Eikon |
| Non-manufacturing purchaser managers' index, United States | monthly | Eikon |
| Industrial production index, EU27 | monthly | Eurostat |
| Tourist arrivals, Spain | monthly | Eurostat |
| Advanced retail sales | monthly | FRED |
| Coincident economic activity index, United States | monthly | FRED |
| Exports of services, United States | monthly | FRED |
| Manufacturers' new orders, United States | monthly | FRED |
| Manufacturers' shipments, United States | monthly | FRED |
| Total air freight, Hong Kong airport | monthly | HKG |
| Container throughput, Hong Kong port | monthly | Hong Kong Ports |
| Industrial production index, Brazil | monthly | IBGE |
| Exports of services, Singapore | quarterly | IMF |
| Industrial production index, China | monthly | NBS |
| Merchandise exports, China | monthly | NBS |
| Business confidence index, Brazil | monthly | OECD |
| Business confidence index, France | monthly | OECD |
| Business confidence index, Germany | monthly | OECD |
| Business confidence index, Japan | monthly | OECD |
| Business confidence index, Netherlands | monthly | OECD |
| Business confidence index, OECD | monthly | OECD |
| Business confidence index, South Korea | monthly | OECD |
| Business confidence index, United Kingdom | monthly | OECD |
| Business confidence index, United States | monthly | OECD |
| Construction index, Canada | monthly | OECD |

| Variable | Frequency | Source |
| --- | --- | --- |
| Consumer confidence index, Germany | monthly | OECD |
| Export volume of goods and services incl. forecasts, Germany | quarterly | OECD |
| Export volume of goods and services incl. forecasts, Japan | quarterly | OECD |
| Export volume of goods and services incl. forecasts, Netherlands | quarterly | OECD |
| Export volume of goods and services incl. forecasts, United States | quarterly | OECD |
| GDP volume incl. forecasts, Germany | quarterly | OECD |
| GDP volume incl. forecasts, Japan | quarterly | OECD |
| GDP volume incl. forecasts, OECD | quarterly | OECD |
| GDP volume incl. forecasts, United Kingdom | quarterly | OECD |
| GDP volume incl. forecasts, United States | quarterly | OECD |
| Industrial production index, Canada | monthly | OECD |
| Industrial production index, France | monthly | OECD |
| Industrial production index, Germany | monthly | OECD |
| Industrial production index, Italy | monthly | OECD |
| Industrial production index, Japan | monthly | OECD |
| Industrial production index, Mexico | monthly | OECD |
| Industrial production index, OECD | monthly | OECD |
| Industrial production index, Russian Federation | monthly | OECD |
| Industrial production index, South Korea | monthly | OECD |
| Industrial production index, Spain | monthly | OECD |
| Industrial production index, United Kingdom | monthly | OECD |
| Industrial production index, United States | monthly | OECD |
| Manufacturing business activity confidence indicator, Germany | monthly | OECD |
| Manufacturing business activity confidence indicator, Italy | monthly | OECD |
| Manufacturing business activity confidence indicator, Netherlands | monthly | OECD |
| Manufacturing business activity confidence indicator, Poland | monthly | OECD |
| Manufacturing business activity confidence indicator, Switzerland | monthly | OECD |
| Manufacturing employment future tendency, Germany | monthly | OECD |
| Manufacturing employment future tendency, Italy | monthly | OECD |
| Manufacturing employment future tendency, Netherlands | monthly | OECD |
| Manufacturing export order books, France | monthly | OECD |
| Manufacturing export order books, Germany | monthly | OECD |
| Manufacturing export order books, Italy | monthly | OECD |
| Manufacturing export order books, Netherlands | monthly | OECD |
| Manufacturing export order books, United Kingdom | monthly | OECD |
| Manufacturing export order books, United States | monthly | OECD |
| Manufacturing order books, France | monthly | OECD |
| Manufacturing order books, Germany | monthly | OECD |
| Manufacturing order books, Italy | monthly | OECD |
| Manufacturing order books, Netherlands | monthly | OECD |
| Manufacturing order books, United Kingdom | monthly | OECD |
| Merchandise exports, Belgium | monthly | OECD |

| Variable | Frequency | Source |
|---|---|---|
| Merchandise exports, BRIICS | monthly | OECD |
| Merchandise exports, Canada | monthly | OECD |
| Merchandise exports, China | monthly | OECD |
| Merchandise exports, France | monthly | OECD |
| Merchandise exports, Germany | monthly | OECD |
| Merchandise exports, India | monthly | OECD |
| Merchandise exports, Italy | monthly | OECD |
| Merchandise exports, Japan | monthly | OECD |
| Merchandise exports, Mexico | monthly | OECD |
| Merchandise exports, Netherlands | monthly | OECD |
| Merchandise exports, OECD | monthly | OECD |
| Merchandise exports, Russian Federation | monthly | OECD |
| Merchandise exports, South Africa | monthly | OECD |
| Merchandise exports, South Korea | monthly | OECD |
| Merchandise exports, United Kingdom | monthly | OECD |
| Merchandise exports, United States | monthly | OECD |
| Retail trade index, values, Canada | monthly | OECD |
| Retail trade index, values, France | monthly | OECD |
| Retail trade index, values, Italy | monthly | OECD |
| Retail trade index, values, Spain | monthly | OECD |
| Retail trade index, values, United States | monthly | OECD |
| Retail trade index, volumes, Brazil | monthly | OECD |
| Retail trade index, volumes, OECD | monthly | OECD |
| Retail trade index, volumes, United States | monthly | OECD |
| Service confidence indicator, France | monthly | OECD |
| Service confidence indicator, Netherlands | monthly | OECD |
| Service employment future tendency, Netherlands | monthly | OECD |
| Container throughput index | monthly | RWI/ISL |
| Merchandise exports, Singapore | monthly | Singapore DOS |
| Total container throughput, Singapore | monthly | Singapore DOS |
| Export volumes, world | quarterly | UNCTAD |
| Exports of services, world | quarterly | UNCTAD |
| Free market commodity price index | monthly | UNCTAD |
| Export prices of manufactures | monthly | WTO |
| Total merchandise exports | quarterly | WTO |